\documentclass[twocolumn]{aastex63}

\usepackage{graphicx} 
\usepackage{float} 
\usepackage{natbib}
\usepackage{amsmath}
\usepackage[T1]{fontenc}
\usepackage[utf8]{inputenc}
\newcommand{\overbar}[1]{\mkern 1.5mu\overline{\mkern-1.5mu#1\mkern-1.5mu}\mkern 1.5mu}

\shorttitle{Super-eccentric Jupiters}
\shortauthors{Jackson et al.}

\begin{document}
\title{Statistical Analysis of the Dearth of Super-eccentric Jupiters in the Kepler Sample}
\author[0000-0002-0323-4828]{Jonathan M. Jackson}
\affil{Astronomy Department and Van Vleck Observatory, Wesleyan University, 96 Foss Hill Drive, Middletown, CT 06459, USA}
\affil{Department of Astronomy \& Astrophysics, Center for Exoplanets and Habitable Worlds, The Pennsylvania State University, University Park, PA 16802, USA}
\author[0000-0001-9677-1296]{Rebekah I. Dawson}
\affil{Department of Astronomy \& Astrophysics, Center for Exoplanets and Habitable Worlds, The Pennsylvania State University, University Park, PA 16802, USA}
\author[0000-0002-9644-8330]{Billy Quarles}
\affil{Department of Physics, Astronomy, Geosciences and Engineering Technology,
Valdosta State University, Valdosta, GA 31698, USA}
\affil{Center for Relativistic Astrophysics, School of Physics, Georgia Institute of Technology, Atlanta, GA 30332, USA}
\author[0000-0002-3610-6953]{Jiayin Dong}
\altaffiliation{Flatiron Research Fellow}
\affil{Department of Astronomy \& Astrophysics, Center for Exoplanets and Habitable Worlds, The Pennsylvania State University, University Park, PA 16802, USA}
\affil{Center for Computational Astrophysics, Flatiron Institute, 162 Fifth Avenue, New York, NY 10010, USA}
\correspondingauthor{Jonathan M. Jackson}
\email{jjackson01@wesleyan.edu}

\begin{abstract}
Hot Jupiters may have formed in situ, or been delivered to their observed short periods through one of two categories of migration mechanisms: disk migration or high-eccentricity migration. If hot Jupiters were delivered by high-eccentricity migration, we would expect to observe some ``super-eccentric'' Jupiters in the process of migrating. We update a prediction for the number of super-eccentric Jupiters we would expect to observe in the Kepler sample if all hot Jupiters migrated through high-eccentricity migration and estimate the true number observed by Kepler. We find that the observations fail to match the prediction from high-eccentricity migration with 94.3\% confidence and show that high-eccentricity migration can account for at most $\sim62\%$ of the hot Jupiters discovered by Kepler.
\end{abstract}

\section{Introduction}\label{intro}
Hot Jupiters are challenging to form on the orbits we observe them on today (e.g., \citealt{Bod00,Raf06}). There are two main families of theories used to explain the hypothesized migration of these planets: (1) migration via interactions with the protoplanetary disk \citep{Gol80,War97,Ali05,Ida08,Bro11,Bar14}, and (2) high-eccentricity migration (HEM) \citep{Wu03}. In the latter case, the migrating planet is excited to large eccentricity by von Zeipel-Lidov-Kozai oscillations \citep{Von10,Koz62,Lid62} via interactions with a binary star \citep{Wu03,Fab07,Nao12} or planetary \citep{Nao11,Lit11} companion, planet-planet scattering \citep{Ras96,For06,For08,Cha08,Jur08,Mat10,Nag11,Bea12,Bol12}, or secular chaos \citep{Wu11}, and then tidal dissipation in the planet shrinks and circularizes its orbit.

If hot Jupiters are being delivered by HEM -- migrating from extremely eccentric long period orbits to shorter periods and intermediate eccentricities to, finally, very short periods and circular orbits (i.e., hot Jupiters) -- we might catch them in the act at high and moderate eccentricities. Planets along HEM `tracks', defined by their final orbital period at zero eccentricity, have approximately constant angular momentum and transit probability. \citet{Soc12} showed that the number of super-eccentric Jupiters with $e>0.9$ along a given migration track can be predicted from the number of moderately-eccentric Jupiters ($0.2 <e < 0.6$) on the same track and that if HEM is the dominant channel for hot Jupiter delivery, the Kepler mission should have detected a handful of these super-eccentric Jupiters. \citet{Daw15} updated that prediction, accounting for Poisson counting uncertainties and incompleteness, and then compared it to the observed Kepler sample using the ``photoeccentric effect'' \citep[e.g.,][]{Daw12J} to estimate the eccentricities of the transiting Jupiters. \citet{Daw15} found a paucity of super-eccentric Jupiters indicating that their results were inconsistent with the prediction from HEM with 96.9\% confidence. Moreover, they calculated a two-sigma statistical upper limit on the fraction of hot Jupiters that could have migrated through the super-eccentric warm Jupiter parameter space of 44\%. Thus, most hot Jupiters must have bypassed this region of parameter space, perhaps by migrating smoothly through disk migration or engaging in HEM with chaotic tides (e.g., \citealt{Mar95a,Mar95b}; see Section \ref{chaos4}).

Here we update the photoeccentric effect calculation from \citet{Daw15} using new Kepler light curve fits (Quarles, in prep.), updated stellar parameters from Gaia Data Release 2 \citep{Ber20}, and hot Jupiters discovered (or re-classified as hot Jupiters) after that study, resulting in a 45\% larger sample. We then compare the predicted number of super-eccentric Jupiters to those observed to produce a posterior for the fraction of hot Jupiters that can be explained by HEM. In Section \ref{prediction}, we discuss the steady-state HEM prediction from \citet{Soc12}, which was subsequently updated by \citet{Daw15}. We discuss the construction of our updated sample of Kepler Jupiters and calculate eccentricities in Section \ref{observations}. In Section \ref{model}, we compare the expected number of super-eccentric Jupiters to those observed in our updated data set. In Section \ref{disc}, we discuss these results in the context of other possible migration mechanisms and, in Section \ref{conc}, we summarize our work and discuss our conclusions.

\section{Prediction for the expected number of super-eccentric Jupiters in the Kepler sample} \label{prediction}

Here we update the prediction \citep{Soc12,Daw15} for the expected number of super-eccentric Jupiters that should be found in the Kepler sample if all the moderately eccentric Jupiters were delivered by a steady-state flow of high-eccentricity migration (HEM). To predict the number of super-eccentric Jupiters ($e>0.9$, $\overbar{N}_{\rm{sup}}$) expected in the Kepler sample, we must assume a population of migrating proto-hot Jupiters being tidally circularized along a track of constant angular momentum defined by $a_{\rm{final}}=a(1-e^2)$ or, equivalently, $P_{\rm{final}}=P(1-e^2)^{3/2}$, where $a$, $e$, and $P$ represent the current semi-major axis, eccentricity, and orbital period of the migrating planet respectively. The number of super-eccentric Jupiters is then related to the number of moderately-eccentric Jupiters ($0.2<e<0.6$, $\overbar{N}_{\rm{mod}}$) along the same track by a constant ratio ($r$) of time spent at super eccentricities to moderate eccentricities. The ratio $r$ (defined below) depends on the assumed functional form (but not absolute normalization) for the circularization timescale. 

\begin{figure*}
\center{\includegraphics{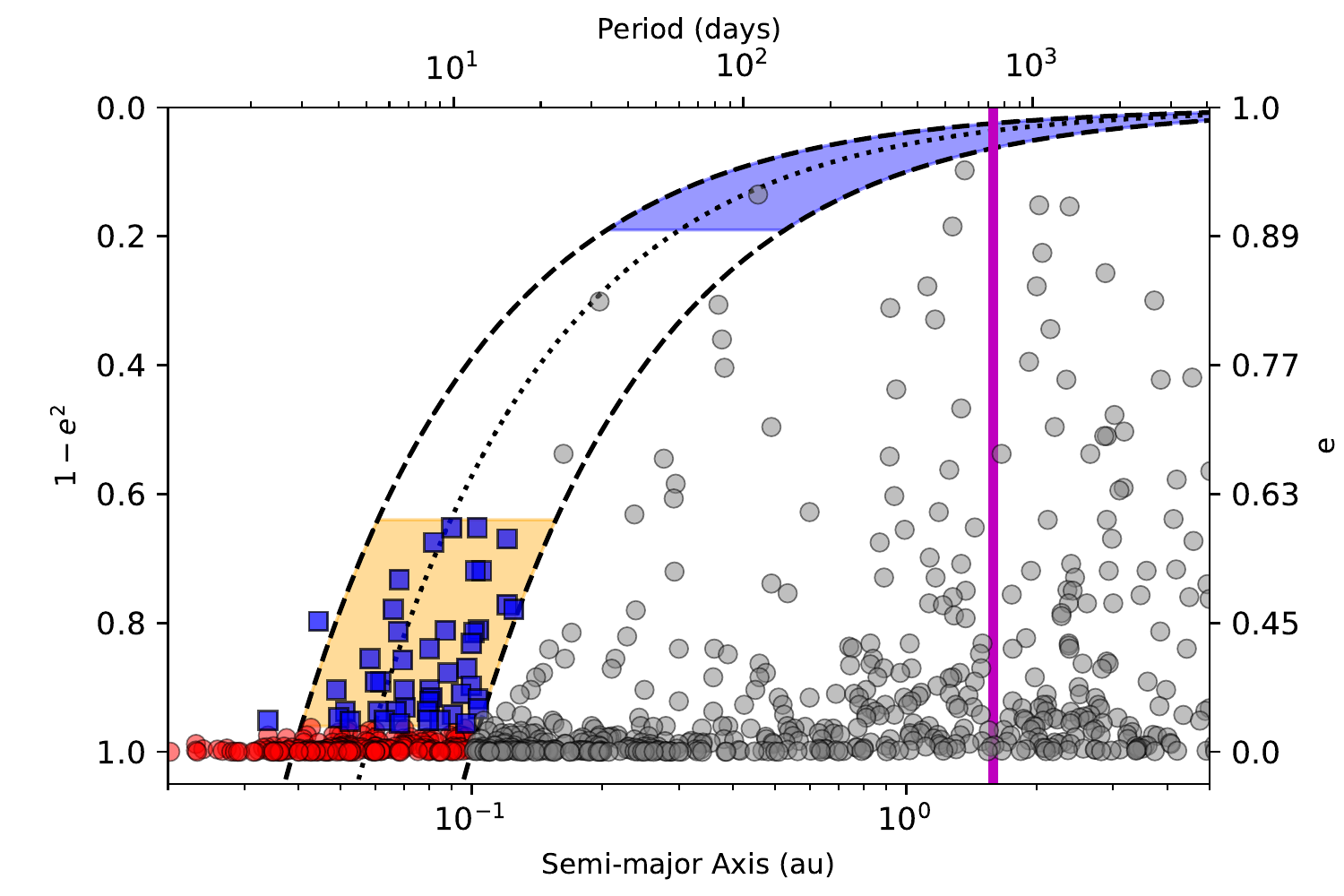}}
\caption
{Eccentricity vs. semi-major axis distribution of Jupiter-mass ($>0.25M_{\rm{Jup}}$) confirmed planets orbiting main sequence stars (see Section \ref{observations} for exact cuts). Here $1-e^2$ is plotted on the y-axis to emphasize elliptical orbits. Hot Jupiters are plotted as red circles, moderately eccentric ($0.2<e<0.6$) Jupiters on tidal migration tracks are plotted as blue squares in the orange-shaded region, and more distant Jupiters ($P>10$ days) are plotted as gray circles. Approximate periods shown on the upper axis are calculated assuming $M_*=1M_{\Sun}$. The range of potential tidal migration tracks is outlined with dashed lines and two intervals of migration tracks, defined by $2.8<P_{\rm{final}}<5$ and $5<P_{\rm{final}}<10$ days, are delineated by the dotted line. The super-eccentric region ($e>0.9$) within these tracks is shaded blue. The upper limit on the completeness of the Kepler sample (2 years) is shown by the vertical purple line. Error bars have been removed for clarity, but see \citet{Daw18} Figure 4 for a similar plot with uncertainties included. Data taken from the NASA Exoplanet Archive on 2022 May 6 \citep{exo_all}.}
\label{fig:ecc4}
\end{figure*}

Since most Kepler Jupiters do not have measured eccentricities, $\overbar{N}_{\rm{mod}}$ is unknown and we must approximate it using a calibration sample of non-Kepler Jupiters, plotted in Figure \ref{fig:ecc4}. We update the calibration sample compiled by \citet{Soc12} and \citet{Daw15} to account for newly discovered planets and updates to their parameters (See Appendix \ref{app:sample}). We include $\overbar{N}_{\rm{mod},0} = 23$ moderately eccentric calibration Jupiters and $\overbar{N}_{P=P_{\rm{final}},0} = 176$ calibration Jupiters with $2.8<P<10$ days, roughly doubling the size of the \citet{Daw15} calibration sample. 

Assuming the fraction of moderately eccentric Jupiters to hot Jupiters is the same between the calibration and Kepler samples, we can then combine this fraction with the number of Kepler Jupiters with $P=P_{\rm{final}}$ ($\overbar{N}_{P=P_{\rm{final}}}$) to calculate the expected number of super-eccentric Jupiters,

\begin{equation} \label{eq:nsup}
    \overbar{N}_{\rm{sup}}=\overbar{N}_{\rm{mod}}r(e_{\rm{max}})=\frac{\overbar{N}_{\rm{mod},0}}{\overbar{N}_{P=P_{\rm{final}},0}}\overbar{N}_{P=P_{\rm{final}}}r(e_{\rm{max}}),
\end{equation}

where $e_{\rm{max}}=[1-(P_{\rm{final}}/P_{\rm{max}})^{2/3}]^{1/2}$ is set by the maximum observable period, $P_{\rm{max}}$. The bar over each $N$ indicates that they are mean numbers. We treat the observationally counted numbers as samples from a Poisson distribution with an unknown mean. For each sample, we assume a Jeffrey's prior for the mean and calculate a posterior distribution of Poisson means, following the procedure outlined in \citet{Daw15} Appendix B. We summarize the means and 95\% confidence intervals of these distributions in Table \ref{tab:sample}.

Accounting for incompleteness, under the tidal time lag approximation \citep{Egg98,Han10,Soc12b,Soc12a,Soc12,Daw15}, the ratio of time spent at super eccentricities to moderate eccentricities for a given $P_{\rm{final}}$ is defined as:

\begin{equation} \label{eq:r}
    r(e_{\rm{max}})=\frac{\int_{0.9}^{e_{\rm{max}}}C_{\rm{comp,sampled}}(e)|\Dot{e}|^{-1}de}{\int_{0.2}^{0.6}|\Dot{e}|^{-1}de},
\end{equation}

where $C_{\rm{comp,sampled}}(e)$ is the completeness accounting for missing data, or the fraction of phases for which we would observe three or more transits during Q1-Q17 of Kepler. See \citet{Daw15} for further discussion of the completeness of the sample.

\begin{figure}
    \centering
    \includegraphics[scale=0.95]{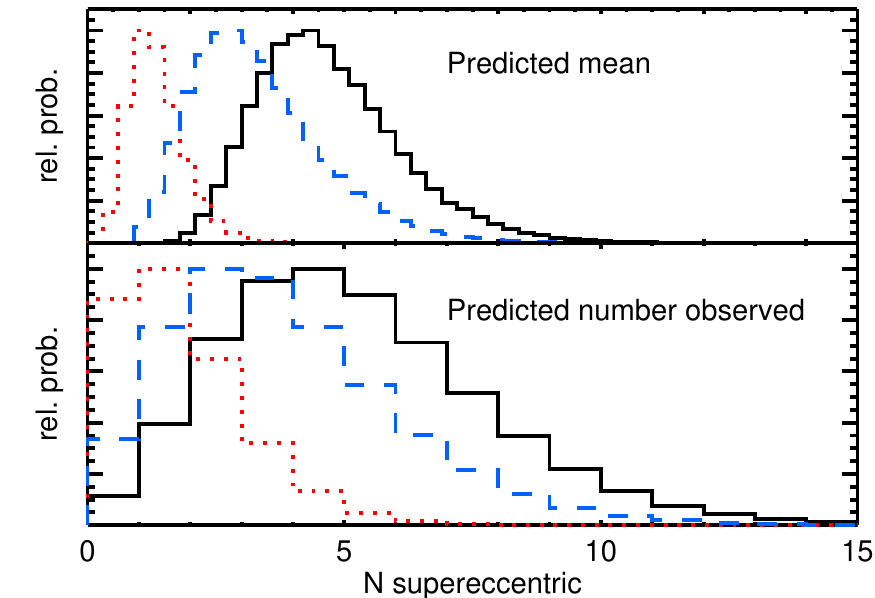}
    \caption
    {Top: Histogram depicting the relative probability distribution of the predicted mean number of super-eccentric Jupiters in $P_{\rm{final}}$ interval 1 (red, dotted), interval 2 (blue, dashed), and combined (black). Bottom: Histogram depicting the relative probability distribution of the expected number of observed super-eccentric Jupiters, created by sampling from the above distribution of Poisson means. }
    \label{fig:pred_hist}
\end{figure}

\citet{Daw15} calculate $r$ to be 0.913, 0.668, 0.422 for $P_{\rm{final}}=$ 2.8, 5, 10 days, respectively. We can plug these values into Equation \ref{eq:nsup} along with our posteriors for $\overbar{N}_{\rm{mod},0}$, $\overbar{N}_{P=P_{\rm{final}},0}$, and $\overbar{N}_{P=P_{\rm{final}}}$ to produce an approximate $\overbar{N}_{\rm{sup}}$. Figure \ref{fig:pred_hist} shows the posteriors for $\overbar{N}_{\rm{sup}}$ in the $P_{\rm{final}}$ ranges 2.8-5, 5-10, and 2.8-10 days. The predicted number observed (bottom panel) is sampled from the distribution of Poisson means (top panel) calculated by Equation \ref{eq:nsup}. With a significantly larger calibration sample than \citet{Daw15}, we predict a similar number super-eccentric Jupiters expected in the Kepler sample with expectation value $\overbar{N}_{\rm{sup}}=4.5^{+1.5}_{-1.1}$. In Section \ref{model}, we will use this prediction for $\overbar{N}_{\rm{sup}}$ in the Kepler sample to set limits on the HEM process.

\begin{deluxetable*}{l|l|lcl|lcl|l}
\tablecolumns{5}
\tablewidth{0pt}
\tablecaption{Counted Planets}
\tablehead{\colhead{$e$} & \colhead{Interval} & \colhead{} & \colhead{Counted} & \colhead{} & \colhead{} & \colhead{Mean\tablenotemark{a}} & \colhead{} & \colhead{Sample\tablenotemark{b}} \\ \colhead{} & \colhead{(days)} & \colhead{} & \colhead{} & \colhead{} & \colhead{} & \colhead{} & \colhead{} &}
\startdata
$0.2<e<0.6$ & 1: 2.8-5 & $N_{\rm{mod},0}$&=&$9$ & $\overbar{N}_{\rm{mod},0}$&=&$9^{+4}_{-3}$ & Cal \\[2pt]
& 2: 5-10 & $N_{\rm{mod},0}$&=&$16$ & $\overbar{N}_{\rm{mod},0}$&=&$16^{+5}_{-4}$ & Cal \\[2pt]
\hline 
unspecified & 1: 2.8-5 & $N_{P=P_{\rm{final}},0}$&=&$134$ & $\overbar{N}_{P=P_{\rm{final}},0}$&=&$134^{+12}_{-11}$ & Cal \\[2pt]
& 1: 2.8-5 & $N_{P=P_{\rm{final}}}$&=&$24$ & $\overbar{N}_{P=P_{\rm{final}}}$&=&$24^{+5}_{-4}$ & Kep \\[2pt]
& 2: 5-10 & $N_{P=P_{\rm{final}},0}$&=&$42$ & $\overbar{N}_{P=P_{\rm{final}},0}$&=&$42^{+7}_{-6}$ & Cal \\[2pt]
& 2: 5-10 & $N_{P=P_{\rm{final}}}$&=&$15$ & $\overbar{N}_{P=P_{\rm{final}}}$&=&$15^{+4}_{-3}$ & Kep \\[2pt]
\enddata
\tablenotetext{a}{Median, with 68.3\% confidence interval, of the posterior of Poisson means, each defining a Poission distribution from which the counted number may be sampled.}
\tablenotetext{b}{Kep: Kepler sample; Cal: calibration non-Kepler sample}
\label{tab:sample}
\end{deluxetable*}

\section{Observations} \label{observations}

Here we update and expand the sample of Kepler warm Jupiter systems from \citet{Daw15} using new stellar parameters from Gaia \citep{Ber20} and improved transit light curve fitting (Quarles, in prep.). These updates improve the precision of the measured stellar densities and light curve fit parameters, which we use to compute probability distributions for each planet candidate's eccentricity.

Following \citet{Daw15}, we identify planet candidates in the Kepler sample with radii ($R_p$) between $8R_{\earth}$ and $22R_{\earth}$ (including candidates whose 95\% radius confidence intervals only slightly overlap with this range), periods ($P$) between 34 days and 2 years, stellar temperatures ($T_{\rm{eff}}$) between $4500K$ and $6500K$, and stellar surface gravities ($\rm{log}g$) $>4$ (or consistent with 4 within the uncertainty). We also require the candidates to transit 3 times in Q1-Q17, have signal-to-noise ratios above 10, and pass basic false-positive vetting according to the Kepler data validation report. Several candidates whose planetary radius posteriors only slightly overlap with our desired range are included, but will be weighted proportionally in our statistical analyses in Section \ref{model}. These cuts produce 45 candidates in our sample, an increase of 45\% from the \citet{Daw15} sample. Many of these initially failed the planet radius cut, but have seen their estimated radii increase with updated stellar parameters from Gaia. Table \ref{tab:candidates} lists all of our candidates and their properties. We note that, although the sample has grown substantially, many of the newly included planet candidates have nearby companions, making them unsuitable for high-eccentricity migration (e.g., \citealt{Mus15}), and are excluded from our calculations in Section \ref{model}, as well as Table \ref{tab:candidates}. We also exclude Kepler-767 b because it has an interior debris disk that would overlap with the planet's orbit if the planet were super eccentric \citep{Law12}. Kepler-419 b, Kepler-539 b, and Kepler 1657 b are also excluded because $e>0.9$ has been ruled out by radial velocity followup and modeling \citep{Daw14a,Man16,Alm18,Jac19,Heb19}. 

We discuss the updates to this sample from the \citet{Daw15} sample in Appendix \ref{app:sample}. We also update the sample of Kepler hot Jupiters, which we describe in Appendix \ref{app:sample}. The net change in the counted number of hot Jupiters, listed in Table \ref{tab:sample}, is 1.

The eccentricity of a planet can affect the observed light curve in many ways and, most notably, can have a detectable effect on the transit duration \citep{Bar07,For08b,Kip08}. We look for highly elliptical planet candidates by combining directly observed parameters from the transit light curve and the star itself. By fitting the light curve under the assumption of a circular orbit, we can produce an estimate of the stellar density under this assumption, $\rho_{\rm{circ}}$, which is related to the true stellar density, $\rho_*$, by:

\begin{equation} \label{eq:photoeccentric}
    \rho_*(e,\omega)g^3(e,\omega)=\rho_{\rm{circ}},
\end{equation}

for which,

\begin{equation} \label{eq:g}
    g(e,\omega)=\frac{1+e\sin{\omega}}{\sqrt{1-e^2}}
\end{equation}

is the approximate ratio of the observed transit speed to the transit speed the planet would have on a circular orbit with the same orbital period for a given argument of pericenter, $\omega$ \citep{Kip10,Daw12J}.

\begin{figure}
\center{\includegraphics[scale=.95]{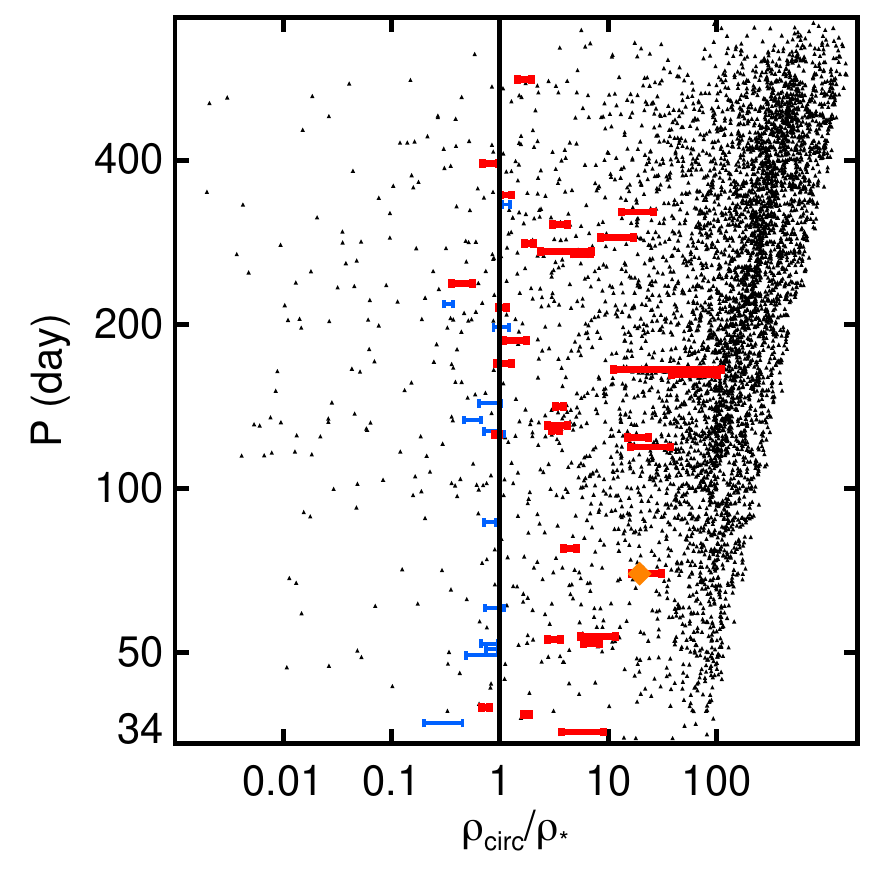}}
\caption{Expected 2D posterior for orbital period $P$ vs. $\rho_{circ}/\rho_*$ of super-eccentric Jupiters. The 45 observed candidates in our sample are overplotted with blue/red bars for planet candidates with/without companions. If the observations matched the prediction, we would expect to see approximately four candidates in the high-probability density region. }
\label{fig:dawson}
\end{figure}

We combine the posteriors of $\rho_{\rm{circ}}$ from Kepler light curves (Quarles, in prep.) and $\rho_*$ from stellar parameters including Gaia measurements \citep{Ber20} into a single $\rho_{\rm{circ}}/\rho_*$ distribution for each of the planets in our sample, marginalized over all other parameters. We then plot the resulting values in Figure \ref{fig:dawson} over the 2-dimensional probability distribution of the predicted super-eccentric warm Jupiters. Red bars represent candidates without known companions and blue bars represent planets with known transiting planet companions. We omit candidates from the plot whose radius posteriors contain $<10\%$ overlap with our warm Jupiter radius constraints. All of the multiplanet systems have interior companions or would result in unstable orbit crossings if the warm Jupiter had an eccentricity $>0.9$.

Based on the prediction in Section \ref{prediction}, we would expect to see $\sim4$-5 candidates in the high-density region of Figure \ref{fig:dawson} if all the moderately eccentric Jupiters were delivered through a steady-state HEM process. While no planets appear to lie directly in the high-probability density region, several planets have some small probability of having $e>0.9$ and $2.8<P_{\rm{final}}<10$. In fact, 6 candidates have mean $\rho_{\rm{circ}}/\rho_*>10$, indicating their eccentricities are unlikely to be zero. We calculate the probability that each candidate is super-eccentric and compare the observations to our predictions in Section \ref{model}.

\begin{deluxetable*}{lllllllll}
\tablecolumns{8}
\tablewidth{0pt}
\tablecaption{Warm Jupiter Planet Candidate Sample}
\tablehead{
\colhead{KIC} & \colhead{KOI} & \colhead{Kepler} & \colhead{Period} & \colhead{$\rho_{\rm{circ}}/\rho_*$\tablenotemark{b}} & \colhead{Radius} & \colhead{Weight\tablenotemark{c}} & \colhead{Estimated $e$\tablenotemark{d}} & \colhead{Probability\tablenotemark{e}} \\[2pt] \colhead{number\tablenotemark{a}} & \colhead{number} & \colhead{number} & \colhead{(days)} & & \colhead{($R_{\earth}$)} & & &
}
\startdata
3534076 & 1209.01 & - & 272.09 & $5.54_{-2.22}^{+4.20}$ & $5.95^{+2.67}_{-0.53}$ & 0.427 & $0.36_{-0.17}^{+0.22}$ & 1.3\%\\[2pt]
3942446 & 1193.01 & - & 119.06 & $22.46_{-5.30}^{+7.20}$ & $10.40^{+2.82}_{-1.89}$ & 0.392 & $0.84_{-0.06}^{+0.09}$ & 7.1\%\\[2pt]
4155328 & 1335.01 & 820b & 127.83 & $2.91_{-0.74}^{+0.47}$ & $9.46^{+1.56}_{-2.14}$ & 0.899 & $0.48_{-0.16}^{+0.30}$ & 4.2\%\\[2pt]
4544670 & 815.01 & - & 35.74 & $5.47_{-1.77}^{+3.52}$ & $7.61^{+1.72}_{-0.91}$ & 0.553 & $0.64_{-0.15}^{+0.20}$ & 0.1\%\\[2pt] 
4566848 & 5071.01 & - & 180.41 & $14.43_{-3.33}^{+5.37}$ & $8.86^{+2.60}_{-0.86}$ & 0.099 & $0.78_{-0.08}^{+0.14}$ & 1.0\%\\[2pt]
4918309 & 1582.01 & - & 186.44 & $1.74_{-0.54}^{+1.62}$ & $11.32^{+1.80}_{-3.35}$ & 0.049 & $0.38_{-0.26}^{+0.34}$ & 0.1\%\\[2pt]
5184584 & 1564.01 & 891b & 53.45 & $9.96_{-2.25}^{+2.67}$ & $5.82^{+1.49}_{-0.50}$ & 1.000 & $0.73_{-0.09}^{+0.14}$ & 4.7\%\\[2pt]
6504954 & 1790.01 & 952b & 130.35 & $5.69_{-0.95}^{+0.76}$ & $7.70^{+2.53}_{-0.84}$ & 0.351 & $0.62_{-0.09}^{+0.21}$ & 1.8\%$^*$\\[2pt]
6766634 & 1375.01 & 1746b & 321.21 & $18.15_{-3.78}^{+6.99}$ & $7.25^{+1.16}_{-1.17}$ & 0.990 & $0.81_{-0.07}^{+0.11}$ & 5.4\%\\[2pt]
7211141 & 1355.01 & 827b & 51.93 & $5.85_{-2.09}^{+1.65}$ & $8.49^{+1.40}_{-1.86}$ & 0.445 & $0.62_{-0.13}^{+0.19}$ & 1.7\%\\[2pt]
7619236 & 682.01 & - & 562.70 & $1.67_{-0.21}^{+0.27}$ & $10.90^{+1.96}_{-1.72}$ & 0.992 & $0.31_{-0.14}^{+0.37}$ & 1.0\%\\[2pt]
7811397 & 1477.01 & - & 169.50 & $1.16_{-0.67}^{+0.53}$ & $9.06^{+0.90}_{-0.28}$ & 0.907 & $0.27_{-0.21}^{+0.35}$ & 1.6\%\\[2pt]
7951018 & 1553.01 & 890b & 52.76 & $3.14_{-0.41}^{+0.48}$ & $8.76^{+2.91}_{-0.97}$ & 1.000 & $0.51_{-0.12}^{+0.29}$ & 2.7\%\\[2pt]
7984047 & 1552.01 & - & 77.63 & $4.56_{-0.67}^{+0.57}$ & $9.40^{+0.67}_{-0.92}$ & 1.000 & $0.58_{-0.11}^{+0.26}$ & 3.8\%\\[2pt]
8672910 & 918.01 & 725b & 39.64 & $0.75_{-0.05}^{+0.06}$ & $10.98^{+0.72}_{-0.99}$ & 1.000 & $0.22_{-0.12}^{+0.41}$ & 0.6\%\\[2pt]
8813698 & 1268.01 & - & 268.94 & $6.14_{-1.01}^{+1.65}$ & $13.57^{+2.30}_{-2.31}$ & 0.998 & $0.66_{-0.11}^{+0.22}$ & 4.5\%\\[2pt]
8827930 & 3801.01 & - & 288.31 & $14.75_{-3.33}^{+3.45}$ & $13.21^{+2.07}_{-2.30}$ & 0.998 & $0.79_{-0.08}^{+0.12}$ & 5.9\%\\[2pt]
9147029 & 5626.01 & - & 397.50 & $0.41_{-0.09}^{+0.45}$ & $6.57^{+3.9}_{-0.91}$ & 0.004 & $0.42_{-0.07}^{+0.0}$ & 0.1\%\\[2pt]
9425139 & 1411.01 & - & 305.08 & $4.17\pm0.82$ & $7.82^{+1.25}_{-0.84}$ & 0.023 & $0.58_{-0.16}^{+0.21}$ & 0.1\%\\[2pt]
9480535 & 3901.01 & 1527b & 160.13 & $0.04_{-0.01}^{+0.03}$ & $6.67^{+1.19}_{-1.47}$ & 0.004 & $0.74_{-0.02}^{+0.20}$ & 0.1\%$^*$\\[2pt]
9512981 & 1466.01 & - & 281.56 & $1.85_{-0.16}^{+0.18}$ & $10.83^{+0.37}_{-0.70}$ & 0.995 & $0.36_{-0.14}^{+0.34}$ & 2.0\%\\[2pt]
9958387 & 2677.01 & - & 237.79 & $0.44_{-0.08}^{+0.09}$ & $9.91^{+1.84}_{-4.28}$ & 0.971 & $0.44_{-0.17}^{+0.31}$ & 2.0\%\\[2pt]
10265602 & 2689.01 & - & 165.34 & $78.73_{-61.44}^{+87.19}$ & $6.98^{+1.53}_{-0.82}$ & 0.901 & $0.85^{+0.11}_{-0.10}$ & 17.3\%$^*$\\[2pt]
10656508 & 211.01 & - & 124.04 & $21.87_{-3.09}^{+3.42}$ & $11.13\pm1.80$ & 1.000 & $0.84_{-0.05}^{+0.09}$ & 14.4\%$^*$\\[2pt]
10795103 & 3683.01 & 1515b & 214.31 & $1.05_{-0.08}^{+0.10}$ & $9.17^{+1.64}_{-0.93}$ & 1.000 & $0.10_{-0.08}^{+0.41}$ & 1.1\%\\[2pt]
11027624 & 1439.01 & 849b & 394.62 & $0.90_{-0.13}^{+0.14}$ & $7.79^{+1.27}_{-1.90}$ & 0.768 & $0.15_{-0.12}^{+0.41}$ & 0.7\%\\[2pt]
11075279 & 1431.01 & - & 345.16 & $1.17_{-0.12}^{+0.15}$ & $7.79^{+1.13}_{-0.36}$ & 0.355 & $0.17_{-0.12}^{+0.35}$ & 0.2\%\\[2pt]
11449844 & 125.01 & 468b & 38.48 & $1.73_{-0.09}^{+0.12}$ & $13.26^{+1.78}_{-0.59}$ & 1.000 & $0.34_{-0.14}^{+0.39}$ & 0.5\%$^*$\\[2pt]
\enddata
\tablenotetext{a}{KIC number of host star}
\tablenotetext{b}{Posterior median and 68\% confidence interval of the ratio of mean $\rho_{\rm{circ}}$ to mean $\rho_*$}
\tablenotetext{c}{Fraction of planetary radius posterior that falls within 8-22$R_{\earth}$}
\tablenotetext{d}{See Section \ref{model}}
\tablenotetext{e}{Posterior probability of $e>0.9$ and $2.8<P_{\rm{final}}<10$ days. Values marked with $^*$ indicate candidates those that appear in the eclipsing binary catalog (\citealt{Kir16}; see Section \ref{mcmc}).}
\label{tab:candidates}
\end{deluxetable*}

\section{Comparison between predictions and observations} \label{model}

While we can visually see in Figure \ref{fig:dawson} that our prediction of $\sim4$-5 super-eccentric Jupiters does not appear to be matched by the observations, we must now calculate the probability that each observed planet candidate lies in the super-eccentric range of parameter space and total these probabilities to determine if the observations fail to  match the prediction with statistical significance. We perform this calculation in Section \ref{mcmc} and use these results to estimate an upper limit on the fraction of hot Jupiters that may have arrived via HEM in Section \ref{upper}.

\subsection{Consistency of the Observations with the Prediction} \label{mcmc}

While none of our candidates appear to lie in the high-probability density region of Figure \ref{fig:dawson}, each has some small chance of truly being super-eccentric ($e>0.9$ and $2.8<P_{\rm{final}}<10$ days). A super-eccentric orbit is a possibility because $\rho_{\rm{circ}}/\rho_*$ is a function of both $e$ and $\omega$ (see Equation \ref{eq:photoeccentric}). For very particular ranges of $\omega$, these candidates could have extreme eccentricities while maintaining  $\rho_{\rm{circ}}/\rho_*$ close to 1. To produce an eccentricity posterior distribution for each planet candidate,we run a Markov Chain Monte Carlo (MCMC) exploration of $\rho_{\rm{circ}}$, $\rho_*$, $e$, and $\omega$, following the procedure outlined in \citet{Daw12J} Section 3.4.

To estimate $N_{\rm{sup}}$, we  conduct a separate Monte Carlo process in which we draw from the eccentricity posteriors of each candidate $10^6$ times. In each trial, we compute $P_{\rm{final}}$ for each candidate and count $N_{\rm{sup}}$, the number of planets with $e>0.9$ and $P_{\rm{final}}$ falling within each of our two intervals (2.8-5 or 5-10 days), weighted by the candidate's overlap with the warm Jupiter radius range. We then draw from the predicted $N_{\rm{sup}}$ posterior from Section \ref{prediction} for each interval. If the observed $N_{\rm{sup}}$ in each interval is greater than or equal to the predicted $N_{\rm{sup}}$ in a given trial, we count that trial as a success. 

\begin{figure}
    \centering
    \includegraphics[scale=.95]{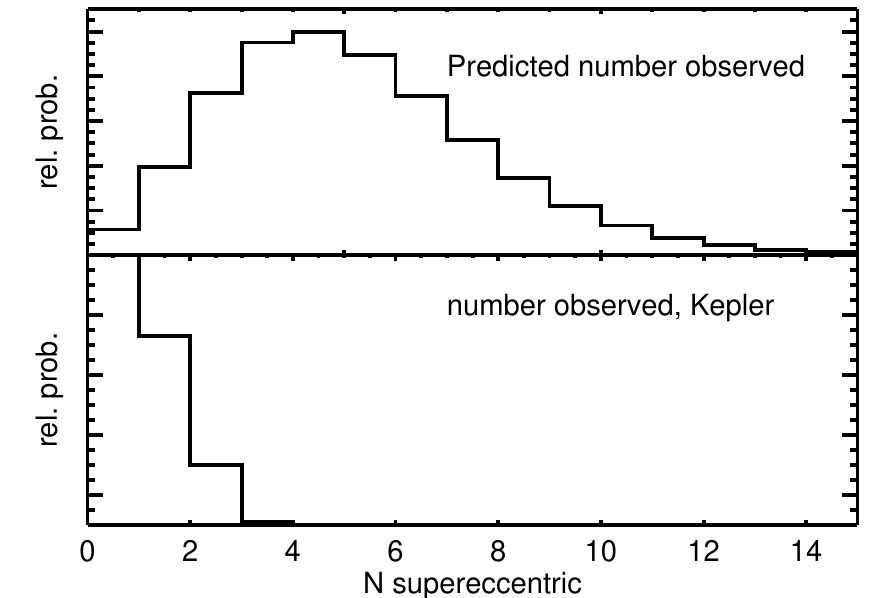}
    \caption{Top: Histogram depicting the relative probability distribution of the expected number of super-eccentric Jupiters from our prediction in Section \ref{prediction} (identical to bottom panel of Figure \ref{fig:pred_hist}). Bottom: histogram depicting the relative probability distribution of super-eccentric Jupiters observed in the Kepler sample. The clear difference between these distributions demonstrates that the observations deviate from the  prediction.}
    \label{fig:new_hist}
\end{figure}

\begin{figure*}
    \centering
    \includegraphics{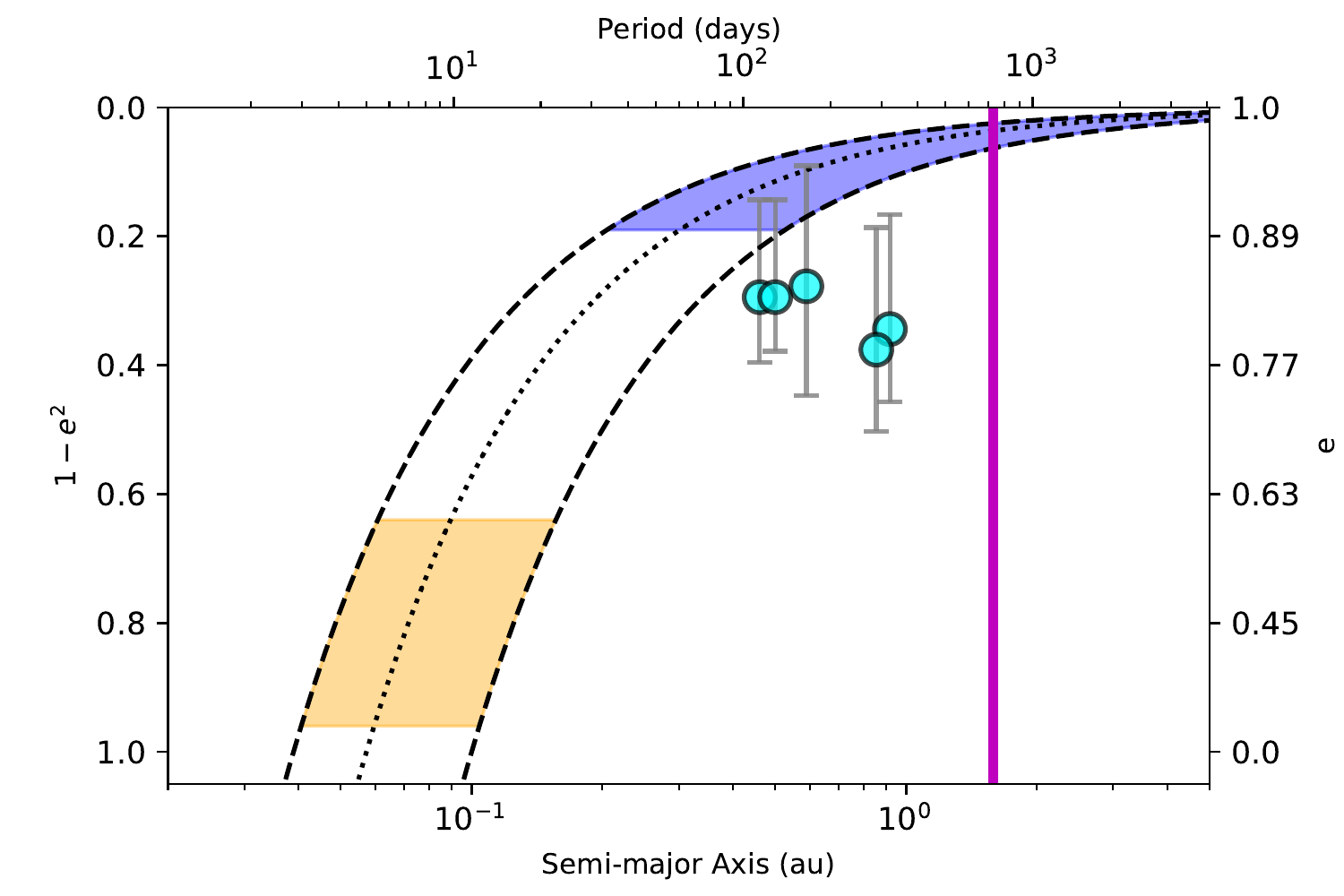}
    \caption{Eccentricity vs. semi-major axis of the 5 planet candidates with the greatest probability of being super-eccentric Jupiters (blue shaded region). From left to right, these are KOI-1193.01, KOI-211.01, KOI-2689.01, KOI3801.01, and Kepler-1746 b. The error bars represent the $68\%$ confidence interval for each candidate's eccentricity. Three of these confidence intervals cross into the super-eccentric region of parameter space. Shaded regions and deliminator lines are the same as in Figure \ref{fig:ecc4}.}
    \label{fig:ecc_super}
\end{figure*}

We measure an $N_{\rm{sup}}$ posterior with a median of $N_{\rm{sup}}=1^{+1}_{-1}$ and find that we detected too few super-eccentric Jupiters to match our prediction with 94.3\% confidence. Figure \ref{fig:new_hist} shows the distribution of counted super-eccentric planets in our trials (bottom panel) compared to the predicted number (top panel) from Figure \ref{fig:pred_hist}. We list the posterior probability that each planet in our sample is super-eccentric in Table \ref{tab:candidates}. The highest probability of super-eccentricity is KOI-2689.01 with $17.3\%$ and only five candidates have probabilities $\ge5\%$. We plot the measured $68\%$ eccentricity confidence interval of each of these five candidates in Figure \ref{fig:ecc_super}. If the true eccentricity of each of these planets lies in the blue shaded region, our prediction would be confirmed by the observations and HEM could not be ruled out as a delivery mechanism for all hot Jupiters.

We note that five candidates in our sample, including KOI-2689.01 and KOI-211.01, are listed in the Kepler eclipsing binary catalog \citep{Kir16}. These candidates may still be planets, but have been identified as presenting features indicative of eclipsing binaries in their transit light curves. Because we cannot confirm these candidates as false positives, we include them in our analysis (as did \citealt{Daw15}) and mark them as potential eclipsing binaries in Table \ref{tab:candidates}. Rerunning our calculations without these systems, we find that we detected too few super-eccentric Jupiters to match our prediction with 95.9\% confidence.

Our result is consistent with that of \citet{Daw15}, but with slightly lower confidence (94.3\% vs. 96.9\%). Despite the higher $\rho_*$ precision of our updated sample, The addition of new candidates with non-zero probability of being super-eccentric into the sample means that the overall probability we are observing too few super-eccentric planets has decreased.

\subsection{Upper limit on hot Jupiter delivery from HEM} \label{upper}

We now set an upper limit on the delivery of hot Jupiters via the HEM process based on our observations of the Kepler sample. To do this, we rewrite Equation \ref{eq:nsup} to include the fraction of hot Jupiters that arrived by HEM, $f_*$. 

\begin{equation} \label{eq:frac}
    \overbar{N}_{\rm{sup}}=f_*\overbar{N}_{\rm{mod}}r(e_{\rm{max}})
\end{equation}

Previously, this equation assumed that all hot Jupiters came from HEM so that we could test the prediction derived from this assumption against the observations. Now we can rerun the Monte Carlo calculation from Section \ref{mcmc} with our new factor included. We implement a modified Jeffrey's prior on $f_*$ following \citet{Daw15} such that it is uniform between 0 and 5\% and scales with $1/f_*$ above $f_*=5\%$. We then draw from the eccentricity posteriors of each candidate $10^6$ times and count the number of super-eccentric Jupiters ($N_{\rm{sup}}$). We consider the trials in which the observed $N_{\rm{sup}}$ meets that drawn from the modified prediction (Equation \ref{eq:frac}) to be successes and derive a final posterior for $f_*$ from these trials.

We find a median of 10\% and a two-sigma upper limit of 62\% on $f_*$, or the fraction of hot Jupiters that could have been delivered by HEM. This limit is larger than that found by \citet{Daw15}, but still implies at least $\sim1/3$ of hot Jupiters must have been delivered by other mechanisms. We discuss caveats to this conclusion in Section \ref{disc}. If the candidates on the eclipsing binary catalog are removed, the two-sigma upper limit on $f_*$ drops to 52\%.

\section{Discussion} \label{disc}

In Section \ref{model}, we showed that HEM, the process by which a highly eccentric giant planet is tidally circularized during close stellar passages resulting in a hot Jupiter, cannot account for all of the observed hot Jupiters. Here we discuss alternatives to HEM (Section \ref{alternatives}), an important caveat to our primary result (Section \ref{planetplanet}), and some ways that migrating planets may be able to move through the super-eccentric phase more quickly than expected (Section \ref{chaos4}).

\subsection{Alternatives to HEM} \label{alternatives}

If hot Jupiters were not delivered by HEM, they must have either (a) formed in situ, (b) migrated via disk-driven migration, or (c) been implanted directly through planet-planet scattering. The final of these is unlikely to account for the majority of hot Jupiters due to the large change in orbital energy required. Both disk migration and in situ formation remain under investigation as plausible origins mechanisms.
 In situ formation of hot Jupiters is challenging due to the large quantity of solid material needed to form a planetary core \citep{Raf06}. In situ formation is also difficult to reconcile with the lack of nearby planets in the majority of hot Jupiter systems \citep{Hua16,Poo21}. Disk migration, in which competing torques in a gas disk may result in a net inward migration of a growing gas giant \citep{Gol80,War97,Ali05,Ida08,Bro11}, has been a common explanation for hot Jupiters. However, some studies have found that disk conditions may not be conducive to inward migration of gas giants (e.g., \citealt{Fun18}). Another challenge with the in situ formation and disk scenarios is that planet-planet scattering \citep{Pet14,And20}
and planet-disk interactions \citep{Duf15} are not able to excite the observed moderate eccentricities so close to the star.

\subsection{The Effect of Planet-planet von Zeipel-Lidov-Kozai Migration} \label{planetplanet}

Warm Jupiters undergoing planet-planet von Zeipel-Lidov-Kozai oscillations leading to HEM can remain unquenched until they reach very short periods if the perturber is nearby \citep{Don14,Jac21}, meaning we will tend to observe them at eccentricities below their tidal migration tracks. This does not affect the total number of super-eccentric Jupiters predicted by Equation \ref{eq:nsup} because that equation uses the ratio of time spent at super-eccentricities to moderate eccentricities while on the track, irrespective of how long a planet spends off of it. However, as \citet{Daw15} argued, it is possible that some of the moderately eccentric planets included in the calibration sample are undergoing von Zeipel-Lidov-Kozai oscillations and tidally migrating on a track with $P_{\rm{final}}<2.8$ days, and thus do not truly belong in the $2.8 < P_{\rm{final}}< 10$ days interval. For some moderately eccentric Jupiters, existing radial-velocity observations can rule out such companions \citep{Don14} -- for example, a Jupiter mass companion would need to be located within 0.6 au to remain coupled to a hot Jupiter located at 0.05 au -- but others need a longer baseline of observations.

However, we note that the timescale for tidal circulation is very strongly dependent on $P_{\rm{final}}$. If migrating Jupiters with $P_{\rm{final}} < 2.8$ days that have not yet circularized are common, it is unlikely that hot Jupiters with periods between 2.8 and 10 days could have circularized over the age of the universe. Thus, we still need some explanation other than HEM to account for these hot Jupiters and any observed moderately eccentric Jupiters not currently undergoing planet-planet von Zeipel-Lidov-Kozai oscillations.

\subsection{Theoretical Methods of Bypassing the Super-eccentric Phase} \label{chaos4}

An alternative explanation for the dearth of super-eccentric Jupiters in the Kepler sample is that hot Jupiters are migrating via HEM, but are bypassing the super-eccentric phase more quickly than that assumed from equilibrium tides. Chaotic or diffusive tides (e.g., \citealt{Mar95a,Mar95b}) may also contribute to the dissipation inside the planet. Recent efforts to model chaotic tides in migrating Jupiters show that, if the large amount of energy produced can be efficiently dissipated in the planet, the HEM process can be sped up in its early stages at high eccentricities \citep{Iva04,Wu18,Vic18,Vic19,Yu21}. However, this mechanism is most effective for planets undergoing periapse passages very close to the star and thus may not have operated for many of the moderately eccentric Jupiters in our sample, particularly those with $P_{\rm{final}}>5$ days.

Another possibility is that migrating super-eccentric Jupiters are thermally inflated due to residual heat from their core-accretion phase that has not yet dissipated. \citet{Roz21} show that Jupiters undergoing ``inflated eccentric migration'' would move through the early stages of HEM more quickly, and predict warm Jupiters with $e>0.9$ to be very rare.

A related explanation is that moderately eccentric hot Jupiters may begin HEM much closer to the star (e.g., within $\sim 0.5 \rm{au}$), following in situ formation or disk migration, never traveling through much of the blue region in Figure \ref{fig:ecc4}.

\section{Conclusions} \label{conc}

\citet{Daw15} showed that there is a paucity of observed super-eccentric Jupiters in the Kepler sample if all or most hot Jupiters are delivered via HEM \citep{Soc12}. We have updated this calculation using improved light curve fits (Quarles, in prep.) and stellar parameters from Gaia \citep{Ber20}. We confirm the paucity of super-eccentric Jupiters in the Kepler sample with 94.3\% confidence and set a 95\% upper limit of 62\% on the fraction of hot Jupiters that could have arrived through HEM, assuming steady tidal circularization with equilibrium tides. Thus, some other mechanism must account for at least $\sim1/3$ of the known hot Jupiters.

In Section \ref{disc}, we discussed some possible alternatives to the standard HEM picture we assumed for our calculations. Hot Jupiters may have formed in situ or been delivered by disk migration. In this case, planet-planet scattering or some other mechanism must be invoked to explain the observed moderately eccentric Jupiters on tidal migration tracks. Another alternative is that most hot Jupiters did arrive through HEM, but bypassed the super-eccentric phase through chaotic tides, inflated eccentric migration, or eccentricity excitation after forming as in situ warm Jupiters or following disk migration.

While the TESS mission will not provide a population of Jupiters complete enough to large periods on which to repeat a similar calculation, it may provide some new eccentric planets to improve our constraints on the minimum contribution of HEM. For example, \citet{Don21b} recently discovered TOI-3362b, a massive planet candidate on a 18.1 day orbital period with $e=0.815^{+0.023}_{-0.032}$, which would be very difficult to explain with planet-planet scattering. Radial velocity follow up of TESS hot and warm Jupiters will help us better understand the eccentricity and multiplicity distributions of these planets, which may improve our constraints on HEM as a hot Jupiter delivery mechanism.

\acknowledgments
We thank the anonymous referee for their helpful comments on this paper. We gratefully acknowledge support from grant NNX16AB50G awarded by the NASA Exoplanets Research Program and the Alfred P. Sloan Foundation's Sloan Research Fellowship. Computations for this research were performed on the Pennsylvania State University’s Institute for Computational and Data Sciences’ Roar supercomputer. This research was supported in part through research cyberinfrastructure resources and services provided by the Partnership for an Advanced Computing Environment (PACE) at the Georgia Institute of Technology. The Center for Exoplanets and Habitable Worlds is supported by the Pennsylvania State University, the Eberly College of Science, and the Pennsylvania Space Grant Consortium. This research has made use of the NASA Exoplanet Archive, which is operated by the California Institute of Technology, under contract with the National Aeronautics and Space Administration under the Exoplanet Exploration Program.

\appendix

\section{Updates to the sample of giant planet candidates} \label{app:sample}

Here we describe our updates to the various samples of giant planet candidates from \citet{Daw15} that we use in our calculations, including the sample of long-period Kepler Jupiters, the sample of short-period Kepler Jupiters in each $p_{\rm{final}}$ interval, the sample of short-period non-Kepler Jupiters in $p_{\rm{final}}$ interval, and the sample of moderately eccentric Jupiters in each $p_{\rm{final}}$ interval. For each sample, we began with every planet candidate that fell within our planetary radius and stellar parameter cuts and checked for false positive signatures including secondary eclipses inconsistent with planethood, large centroid-transit correlations, different odd/even eclipse depths, strong centroid-flux correlations, and large position offsets from the KIC catalog. Below are the additions to and removals from the \citet{Daw15} sample using our updated knowledge of the candidates. All totals excluding the long-period Jupiters are listed in Table \ref{tab:sample}.
\begin{enumerate}
    \item Long-period giant planet candidates: We exclude KOI-1483.01 and KOI-5241.01, which appeared in the \citet{Daw15} sample, because they have since been identified as false positives \citep{Mor16}. We also exclude KOI-458.01, which orbits an evolved star, and KOI-1587.01, whose radius has been revised and is now too large \citep{Ber20}. We exclude KIC 9413313 because it does not appear in the \citet{Ber20} catalog or on the NASA Exoplanet Archive. All of our additions to the \citep{Daw15} sample had estimated radii below $8R_{\earth}$ but were revised to fall within our radius cut by \citet{Ber20}.
    \item Moderately eccentric Jupiters ($N_{\rm{mod},0}$) with $2.8<P_{\rm{final}}<5$ days: We add WASP-150 b, WASP-186 b, and NGC 2682 YBP 1514 b to the sample from \citet{Daw15}, which have all been discovered since its publication, for a net increase of 3 candidates.
    \item Moderately eccentric Jupiters ($N_{\rm{mod},0}$) with $5<P_{\rm{final}}<10$ days: We add CoRoT-20 b, TIC 237913194 b, TOI-172 b, TOI-677 b, WASP-117 b, WASP-148 b, WASP-162 b, WASP 185 b, and NGC 2682 YBP 1194 b to the sample from \citet{Daw15}, which have all been discovered since its publication or have had their eccentricities constrained, for a net increase of 9 candidates.
    \item Non-Kepler Jupiters with $2.8<P<5$ days ($N_{P=P_{\rm{final}},0}$): We remove HAT-P-35 b from the \citet{Daw15} sample because its radius estimate has changed with new Gaia data and we remove OGLE-TR-211 b because its most recent eccentricity estimate is ambiguous. We add CoRoT-27 b, CoRoT-29 b, HAT-P-3 b, HAT-P-42 b, HAT-P-44 b, HAT-P-45 b, HAT-P-50 b, HAT-P-51 b, HAT-P-55 b, HAT-P-58 b, HAT-P-59 b, HAT-P-63 b, HAT-P-64 b, HAT-P-9 b, HATS-12 b, HATS-13 b, HATS-22 b, HATS-25 b, HATS-29 b, HATS-30 b, HATS-36 b, HATS-47 b, HATS-48 A b, HATS-5 b, HATS-55 b, HATS-60 b, HATS-63 b, HATS-65 b, HATS-68 b, KELT-8 b, NGTS-13 b, TOI-1296 b, TOI-1298 b, TOI-628 b, TOI-905 b, TrES-1 b, TrES-4 b, WASP-101 b, WASP-120 b, WASP-124 b, WASP-126 b, WASP-13 b, WASP-153 b, WASP-20 b, WASP-28 b, WASP-39 b, WASP-53 b, WASP-56 b, WASP-57 b, WASP-60 b, WASP-69 b, WASP-70 A b, WASP-83 b, WASP-94 A b, WASP-96 b, WASP-98 b, WTS-1 b, XO-1 b, XO-4 b, XO-5 b, XO-7 b, K2-29 b, K2-30 b, K2-34 b, HD 103720 b, and HD 202772 A b, which have all been discovered since 2015 or have had their eccentricities constrained, for a net increase of 65 candidates.
    \item Non-Kepler Jupiters with $5<P<10$ days ($N_{P=P_{\rm{final}},0}$): We add CoRoT-20 b, CoRoT-30 b, HATS-61 b, HATS-72 b, KELT-6 b, TOI-1268 b, TOI-172 b, TOI-559 b, WASP-106 b, WASP-129 b, WASP-148 b, WASP-150 b, WASP-162 b, WASP-185 b, WASP-186 b, WASP-58 b, WASP-68 b, WASP-84 b, WASP-99 b, HD 103774 b, HD 168746 b, HD 285507 b, HIP 91258 b, NGC 2682 YBP 1194 b, and NGC 2682 YBP 1514 b to the sample from \citet{Daw15}, which have all been discovered since its publication or have had their eccentricities constrained, for a net increase of 24 candidates.
    \item Kepler Jupiters with $2.8<P<5$ days ($N_{P=P_{\rm{final}}}$): We add Kepler-1935 b to the sample from \citet{Daw15}, whose radius posterior now overlaps with our target range \citep{Ber20}. We also remove KOI-838.01, which has been identified as a false positive, and are left with a total $N_{P=P_{\rm{final}}}$ of 24 for a net change of 0.
    \item Kepler Jupiters with $5<P<10$ days ($N_{P=P_{\rm{final}}}$): We add Kepler-677 b to the sample from \citet{Daw15}, whose radius posterior now overlaps with our target range \citep{Ber20}. We also remove KOI-774.01 and KOI-1391.01, which have been identified as false positives, and are left with a total $N_{P=P_{\rm{final}}}$ of 15 for a net change of 1 fewer candidate.
\end{enumerate}

\bibliography{astrobib}

\end{document}